# Non procedural language for parallel programs

R. Nuriyev (renat.nuriyev@gmail.com)

**Abstract**

Probably building non procedural languages is the most prospective way for parallel programming just because non procedural means no fixed way for execution. The article consists of 3 parts. In first part we consider formal systems for definition a named datasets and studying an expression power of different subclasses. In the second part we consider a complexity of algorithms of building sets by the definitions. In third part we consider a fullness and flexibility of the class of program based data set definitions.

**Introduction**

Probably building non procedural languages is the most prospective way for parallel programming just because non procedural means no fixed way for execution: we are free to select paths what are better fits the case. Later we will see that finding a way of execution for a given nonprocedural description might be a heavy task as well.

Parallel programs may be considered as a process of transformation some data sets to other data sets on each step. And such view is already in practice. For example, airline company decision "ticket price for passengers travel more than 5 000 miles per year have to be 90% of original price" is based on definition and transformation of set of tickets of passengers with some property. A sequel programs are just a case when these sets have a fixed number of elements.

There are two major ways for definition a set of data: procedural and non-procedural. Procedural way means to present some algorithm for data selection to the set. Non-procedural way means to define a set without description how to create it.

Rational database languages are an example of the last way. But the languages are not the most powerful. The following is a query that can't be expressed in the RDB languages: **select a set of people sympathetic to each other**. This definition can't be used in a database - it requires a named data sets. It also does not allowed in mathematics because of following "nice" set paradox.

Set X is called good if $\neg X \in X$. If U is a set of all "nice" sets is U nice or not? Both possibilities are ended with contradictions.
If U is "nice" then by definition $\neg(U \in U)$ is true. But U contains all "nice" sets and therefore it must include U itself and $U \in U$ is true. We got a contradiction to our assumption.
If U is not "nice" set then $U \in U$ is true. But U consists of such sets X that $\neg(X \in X)$, so $\neg(U \in U)$ should be true for U. Contradiction.

Here we are studying different forms of data set definitions, the role of names and how they influence to expression power, independency of the forms (if some of them may be eliminated without decreasing a expression power). Also we define a class of programs based on these definitions, determine a functional fullness of this class of programs and flexibility of their data event controls.

## 1. Definitions

**Def . 1.** We call **date** pairs **(d, v)**; **d** is called an **attribute** and **v** is called a **value**. Elements **d** and **v** are taken from some countable sets D and V. They although may be the same dimension Cartesian production too: $\mathbf{d=d_1,\ldots d_n}$ and $\mathbf{v=v_1,\ldots, v_n}$.

Definitions of data sets used here are constructed hierarchically from elementary definitions taken from mathematics and computer science practice. We recognize four types of them:
- **enumeration**, when we point out elements of the set;

- pointing **properties** of set's elements in relation to the elements of the same or other sets;

- using **induction** when we fix the base set of elements and induction predicate **p** (rule) by which we are saying that if an element is in this relation p with already established elements, then this element belongs to the set too;

- defining a set of elements as a **functional image** of elements of other sets.

Each definition also determines the name of the set or names of sets if we are using parameterized naming. So to define a single set we generally have to use a system of definitions with some sets of names.

**Def . 2.** Sub systems of definitions that are using only sets defined in the same system are called *closed* (and *open* otherwise). Set names defined in the system is also called *internal*, all others – *external*.

To be a fully defined, system with external names has to have, besides of start data and interpretation of functional and predicate symbols, some start or *a priory* family of sets. Then a system of definitions is a function from *a priory* family of sets to the family of sets defined by this system.

This view gives us a way to compare different subclasses of definitions by comparing classes of associated operators.

**Def . 3.** We say that one **class of definition is more powerful than another** if class of associated operators includes the class of operators associated with another class of definitions.

We will show that no one of definition types might be eliminated without decreasing expression power. Also expressive power might be increased by extending signature with interpreted predicate
$$p(x,S)\equiv(x\in S).$$

In paragraph 6 we will consider an algorithmic complexity of building sets satisfying systems of the definitions from different classes.

## 2. Subclasses of sets definitions

**Def. 4.** Let N to be a set of natural numbers. **Alphabet** for system of definition is a union of punctuation symbols, parenthesis, special symbols $\forall, \exists, \equiv, \in, \cup, \cap, +, -$ and non-intersected sets P, C, S, F, V, where
$P = \{p_j^i | i \in N, j \in N\}$
$C = \{c_i | i \in N\}$
$S = \{S^i | i \in N\}$
$F = \{f_j^i | i \in N\}$
$V = \{x_i, y_i, z_i, u_i, v_i | i \in N\}$.

Elements of P are called **symbols of predicates**; elements of C are called a **symbols of constants**; elements of F – **symbols of functions**; S – **types of set names**; V – **variables.**

**Def. 5.** Set of **terms** T is defined by induction:
a. empty term $\lambda$ is a term,
b. elements of V and C are terms,
c. if $t_1, \ldots, t_k$ are terms and $f_j^k \in F$, then $f_j^k(t_1, \ldots, t_k)$ is a term.

For term **t** denote v(t) the transitive closed set of variables of t or, more formal, v(t) is a minimal set with next 3 properties:
a. if $t \in V$, then v(t)=t;
b. if $t=f(t_1,\ldots,t_k)$, then $v(t)=v(t_1) \cup \ldots \cup v(t_k)$;
c. if $t \in \{\lambda\} \cup C$, then $v(t) = \emptyset$.

**Def. 6.** An element of $\{S\} \times T \times N$ is called **schema of set names** and is denoted $S_j^t$ ($t \in T$, $j \in N$) and is called a type of name.

**Def. 7.** Expressions $x \in S_j^i$ or $S_j^i \in S_y^m$, where $x \in V$, $y \in T$, are called **elementary selectors** of variable x or of set names $S_y^j$ of type j.

**Def. 8.** A finite set U of elementary selectors $x_1 \in S^{i1}_{t1}, \ldots, x_n \in S^{in}_{tn}$ with all different $x_1, \ldots, x_n$ is called a **selector level** if $(v(t_1) \cup \ldots \cup v(t_n)) \cap \{x_1, \ldots, x_n\} = 0$. The set $(v(t_1) \cup \ldots \cup v(t_n))$ is denoted $U^-$, $\{x_1, \ldots, x_n\}$ is denoted $U^+$.

**Def. 9.** Let U be a finite set of selectors $U_1, \ldots, U_n$. Then U is called **hierarchy of selectors** if for each $x \in U_i^-$ also true $x \in \cup^k_{(i<j)} U_j^+$ and $\forall i \forall j (U_i^+ \cap U_j^+ = \emptyset)$.

A hierarchy of selectors is called **closed** if $U_n=\emptyset$.

Other word, a hierarchy of selectors defines domains for each variable of the form.

**Def. 10. Elementary form** of subset data definitions (**form of definition** for short) is called one of the next expressions:

α. $S_c^i = \{c_1,\ldots,c_k\}$,

β. $S_t^i = \{Q_1x_1 \in S_{t1}\ldots Q_nx_n \in S_{tn}\ p(x_1,..,x_n,z_1,\ldots z_m); U\}$,

γ. $S_t^i = \{S_{t0}^{i0},\ Q_1x_1 \in S_{t1}\ldots Q_nx_n \in S_{tn} p(x,y,x_1,\ldots,x_n, z_1,\ldots,z_m);\ U\}$,

δ. $S_t^i = \{t_0(z_1,\ldots, z_m), U\}$,

where $S_t^i, S_c^i \in S \times T \times N$, $Q_1,\ldots Q_n \in \{\forall, \exists\}$, $p \in P$, $\{z_1,\ldots,z_m\} \subset v(t)$, U is a closed hierarchical selector for variables from $v(t) \cup \{z_1,\ldots,z_m\}$; $c_1,\ldots,c_k \subset C \cup \{S \times C\}$, $x_1,\ldots,x_n, z_1,\ldots,z_m \in W\{S \times W\}$, $t \in T \cup \{S \times T\}$, $v(t) = \cup U_i^+$.

Variables from $v(t)$ are called **parameters** for this form.

Idea behind the forms is as follows.

-Form of type α defines set with name $S_c^i$ and elements $c_1,\ldots, c_k$.

-Form of type β defines family of sets when variables of the term t (names) are taken from sets in selector U and each set is defined by relation **p** its elements to elements of the same or another sets $S_{t1}, \ldots, S_{tn}$ expressed with the quantified formula

$Q_1x_1\ldots Q_nx_n\ p(x_1,.., x_n, z_1,\ldots, z_m)$

and $x_1,.., x_n$ are taken from sets with names $S_{t1,\ldots,} S_{tn}$ correspondently.

**Example**. List of products Pr is an example of set consisting of elements and set names – product may include of another products. The form

$S^{india} = \{\forall S_x \in S^{india}\ \forall y \in S_x\ p(y)\}$,

where p(y) means product **y** is produced in India, defines the set of products completely produced in India.

-Form of type γ defines sets with names $S_t^i$ when variables of term t are selected with selector U. Each set $S_t^i$ is defined by induction and $S_{t0}^{i0}$ is a base of induction, formula $Q_1x_1 \in S_{t1}\ldots Q_nx_n \in S_{tn}p(x,y,x_1,\ldots,x_n, z_1,\ldots,z_m)$ is an induction step – if $y \in S_t^i$ then any element x from universe also belong to $S_t^i$ if this formula is hold.

Example of γ type definition. Let G be a graph $(V,R)$, V is vertices, R is a set of edges. Then the form

$S = \{S_0, p(x,y), y \in V\}$,

where $p(x,y)$ is interpreted as edge $(x, y) \in R$, defines a minimum connected components of the graph G which includes vertices of $S_0$.

Form of type δ defines, or probably better to say creates, a "secondary" data. This form determines sets $S_{vt}^i$ where vt is a value of term **t** when its variables selected with selector U. Elements of set $S_{vt}^i$ are values of the term $t_0$ with arguments selected with selector U. It may be elements out of start data. After applying the definition new data are included in data universe.

Generally, forms are based on another definition forms. So to define a set we have to talk about a system of definitions.

**Def. 11.** A finite set **F** of forms are called a **system of forms** if:
a. left parts of forms are different types;
b. for each set name S of type j the system **F** has a definition of that type name.

### 3. Interpretations of systems of set definitions

**Def. 12.** Set of symbols of predicates, functions, variables and constants used in forms of a system F is called a **full signature** of **F** and **denoted** $\sigma_F$.

**Def. 13.** Let $V = D \times V$ be a **data universe**. To define an **interpretation J** for system of set definitions **F** means to define **start data** $\Omega$ – a finite set of data, finite set of secondary data $\Omega^\lambda \subset \cup f_i^k[V^k]$, and to map
- each **symbol of constant** to element of $\Omega \cup \Omega^\lambda$;
- each **predicate symbol** $p_i^j \in \sigma_F$ to map $[V^j \to \{true, false\}]$
- each **functional symbol** $f_i^j \in \sigma_F$ to map $[V^j \to V]$.

**Terms** used in a form are substitutions of function symbols and variables.

For a given interpretation **J**, set names S(**F**, **J**) also get a concrete names:
$S^{conc}(F, J) = \cup\{S_v^i | v = t(v_1, \ldots, v_n), \{v_1, \ldots, v_n\} \subset \Omega \cup \Omega^\lambda; U\}$, where $S_v^i$ are any schema of names in F and U is a selectors for variables of $\{v_1, \ldots, v_n\}$.

For a given system F and interpretation J we will also use expressions $\Omega = \Omega(J, F)$, $\Omega^\lambda = \Omega^\lambda(J, F)$, $D = D(J, F)$.

### 4. Named data subsets defined with a system of definitions

Let J be an interpretation of a system F, **M** is a set of pairs $(S_v^i, M_v^i)$, where $S_v^i \in S(F,J)$, $M_v^i \subset \Omega \cup \Omega^\lambda \cup S(F, J)$.

**Def. 14.** Let's denote S(**M**) a set of names from pairs of **M**, denote $M(S_v^i, J, F)$ a data set with name $S_v^i$. We will omit F and J if it is clear from context what they are. Second component $M_v^i$ of pair $(S_v^i, M_v^i) \in M$ is called a set **named** $S_v^i$, **M** is called a **named family of sets**.

For a given M and a hierarchical selector U, the map
$\xi \in [\{x1, \ldots, xn\} \to \Omega \cup \Omega^\lambda]$
is called **acceptable**, if for each elementary selection $x \in S_{t(y1,\ldots,yn)}^i$ or $S_x^i \in S_{t(y1,\ldots,yn)}^i$ the relation
$\xi(x) \in M(S_{t(\xi(y1,\ldots,ym))}^i)$
or

$S^i_{\xi(x)} \in M(S^i_{t(\xi(y1,\ldots,ym))})$
is true.

Other word, the map $\xi$ determines the values for variables that satisfy to the selector U.

For a form F with $S^i_{t(y1,\ldots,yn)}$ in a left side and selector U let's denote $\Sigma^F_v$ all acceptable map $\xi$ that $t(\xi(y_1,\ldots,y_n))=v$.

**Remark 1.** The set $\Sigma^F_v$ consists of all data used for a selection data for set $S^i_v$.

The notation $\Sigma_{v|y1,\ldots,yn}$ is used for a set of values for $y_1,\ldots,y_n$.

**Def. 15.** The **value for expression** $Q_1x_1 \in S_{t1} \ldots Q_nx_n \in S_{tn} p(x,y,x_1,\ldots,x_n, z_1,\ldots,z_m)$ on $\Sigma^F_v$ defined by induction on number of quantified variables:

**Induction base**: for formula $Q1x1 \in S_{t1}\ p(y_1,\ldots,y_n,x_1)$ value is a value of non quantified formula
$p(y_1,\ldots,y_n, b_1)*\ldots *p(y_1,\ldots,y_n, b_m)$, where '*' is '∧' if $Q1 = \forall$ or '*' is '∨' if $Q1 = \exists$, $\{b_1,\ldots,b_n\} = \Sigma^F_v$.

**Induction step**: a formula $Q_1x_1 \in S_{t1}\ldots Q_nx_n \in S_{tn} p(x_1,\ldots,x_n, z_1,\ldots,z_m)$ is equal to
$Q_1x_1 \in S_{t1}\ldots Q_nx_n \in S_{tn} p(x_1,\ldots,x_{n-1},b_1, z_1,\ldots,z_m)*\ldots * Q_1x_1 \in S_{t1}\ldots Q_nx_n \in S_{tn} p(x_1,\ldots,x_{n-1},b_m, z_1,\ldots,z_m)$
where * and $b_1,\ldots,b_m$ are defined earlier.

Let's denote as $t[M]$ the set of values of term t on variables values from M.

**Def. 16.** Family M is called **in agreement** with system F on interpretation J, if for each form with left part $S^i_t$ next are true:

a. if F is of type of $\alpha$ and $M^i_c = \{c_1,\ldots,c_k\}$, then $(S^i_c, M^i_c) \in M$,

b. if F is of type of $\beta$, then $(S^i_v, M^i_v) \in M$ if and only if $v \in t[\Omega \cup \Omega^\lambda]$, $\Sigma^F_v \neq \emptyset$ and formula of the form is true on $\Sigma^F_v$,

c. if F is of type $\gamma$, then $(S^i_v, M^i_v) \in M$ if and only if $v \in t[\Omega \cup \Omega^\lambda]$, $\Sigma^F_v \neq \emptyset$, $M^i_v$ is a minimum set with next properties:
   - each element of $M(S^{t(d1,\ldots,dn)}_i)$ belongs to $M(S^i_v)$, where $d_1,\ldots,d_n$ are value of parameters of t for $t(d_1,\ldots,d_n)=v$,
   - if for some $d \in S^i_v$ the formula
   $Q_1x_1 \in S_{t1}\ldots Q_nx_n \in S_{tn} p(d0,d,x_1,\ldots,x_n)$
   is true on $\Sigma^F_v$ for some $d_0$, then $d_0 \in M(S^i_v)$,
   - $M^i_v$ does not have other elements.

d. if F is type of $\gamma$, then $(S^i_v, M^i_v) \in M$ if and only if $v \in t[\Omega \cup \Omega^\lambda]$, $\Sigma^F_v \neq \emptyset$, $M^i_v = [\Omega \cup \Omega^\lambda \cup S(F,J)] \cap t_1[\Sigma^F_{v|v(t1)}]$.

**Def. 17.** A family **M** is called **selected** by a system F on interpretation J, if M is agreed with F and any family $M_1$ have gotten from M by increasing only one of it sets is not agreed with F. A family of named set also is called **a variant of selection.**

**Example 1.** Let us have a list of laboratories and its employee's names, years of experience and salaries. These lists may be defined with forms of type α:

$S_c = (lab_1, \ldots, lab_k)$, // list of labs
$S_{lab\ 1} = \{c^1_1, \ldots, c^1_{n1};\ lab_1 \in S_c\}$, // employees of $lab_1$
........
$S_{lab\ k} = \{c^k_1, \ldots, c^k_{n1};\ lab_k \in S_c\}$ // employees of $lab_k$

Then next form of type β defines the set of names of "nice" laboratories where salary is a monotone growing function of experience:

$S = \{\forall S_{lab} \in S \forall x \in S_{lab}((experience(x) > experience(y)) => (salary(x) > salary(y)),\ S_{lab} \in S_c\}$.

## 5. A comparison of subclasses of forms

Basic questions for system of forms are expressive powers of different sub classes. For comparison of expressiveness we will consider systems of definitions as operators from data universe to the family of named sets. The idea is that if for any system of definitions F from a class *A* exists a system with extended signature from another class *B* that generates for any interpretation J the same family of named sets then we may say *B* is not **weaker** than *A*.

We allow extending signature of forms of class B with:
- new variables,
- **secondary predicates** – logical expressions with ∧¬ ∨ operations with symbols of predicates of a form of class A. Because any logical function can be created with these operations, these logical operations may be replaced with any others.

More formal.
**Def. 18.** Let $\psi^{n1}_1, \ldots, \psi^{nh}_h$ be a function symbols with interpretation $[\{true, false\}^{mi} \rightarrow \{true, false\}]$. Let $p_1, \ldots, p_m$ are predicates symbols. Then **predicate expressions or secondary predicates** we will call such a minimal set of expressions that:
  a. base predicates symbols with arguments of terms $t_1, \ldots, t_k$ build above functional symbols and elements of W, are a predicate expressions;
  b. if ψ is a secondary function symbol and $p_1, \ldots, p_m$ are predicate expressions, then
        $\psi(p_1, \ldots, p_m)$
is a predicate expression as well.

The example of secondary predicates is $p1 \lor ((\neg p1(t1) \land p2(t2)) \lor \neg p3(t3))$, where p1, p2, p3 are base predicates symbols.

From now and up we will consider systems of form, which predicates could be predicate expressions.

**Def. 19. Interpretation** J2 is called **similar** to interpretation J1 if $\Omega(J1)=\Omega(J2)$, $\Omega^\lambda(J1)=\Omega^\lambda(J2)$ and interpretations of their base predicate and functional symbols are the same.
Other word, in similar schemas difference might be in predicate expressions only.

**Def. 20.** Two systems $F_1$ and $F_2$ with same basic predicates and functions symbols are called **equivalent** regarding to name $S_a$, if for any interpretation J for $F_1$ exists a similar interpretation for $F_2$ that sets with name $S_a$ are the same.

**Def. 21.** Let $\Phi_1$ and $\Phi_2$ are subclasses of definitions. Then **$\Phi_1$ is called not less expressive than $\Phi_2$**, notation is $\Phi_1 \geq \Phi_2$, if for each system $F_1 \in \Phi_1$ and set name $S_a$ exists system $F_2 \in \Phi_2$ equivalent to F1 regarding to $S_a$.
The fact that $\Phi_1 \geq \Phi_2$ but not $\Phi_2 = \Phi_1$, will be denoted as $\Phi_1 > \Phi_2$.

Now we will compare next 6 classes:
$G^c$ is a subclass of systems with formulas containing two variables x and y so that $x \in S_y$ belongs to selector of some form;
$G^{\neg \eta}$ is a subclass of systems without forms of type $\eta$ ($\eta \in \{\alpha, \beta, \gamma, \delta\}$);
$G^s$ is a subclass of systems with a signature extended with predicates of $p(x,S) \equiv s \in S$.

The main tool for comparison of expression powers of subclasses are special interpretations similar to free interpretations used in [Luckham D.C., Park D.M., Paterson M.S. On formalized Computer programs. //J. Comp. and Syst. Science. 1970, v4].

**Def 22.** Let F be a system, $\sigma_F^b$ - a set of its functions and predicates. **The standard interpretation** of $\sigma_F^b$ is a pair $(Q \cup Q^\lambda, D)$, where
- Q is a finite set of elements of $q_i$, $i \in N$, called **bearer**;
- $Q^\lambda$ is a finite set of terms build of functional symbols of $\sigma_F^b$ with arguments of Q;
- a diagram D is a finite set of strings "$p_k(t_1,\ldots, t_k)$" or "$\neg p_k(t_1,\ldots, t_k)$" where p is a predicate symbol, $\{t_1,\ldots, t_k\} \subset Q \cup Q^\lambda$, and if "$p_k(t_1,\ldots, t_k)$" is in D then "$\neg p_k(t_1,\ldots, t_k)$" is not in D and inverse, if "$\neg p_k(t_1,\ldots, t_k)$" is in D then "$p_k(t_1,\ldots, t_k)$" is not in D.
- the value of function $f^k \in \sigma_F^b$ on $\{a_1,\ldots, a_k\} \subset Q \cup Q^\lambda$ is an string "$f(a_1,\ldots, a_k)$".
- the value of predicate $p^k \in \sigma_F^b$ is next:

$$p(t_1,\ldots,t_k) = \begin{cases} \text{true, if the string "}p(t_1,\ldots,t_k)\text{" is in D,} \\ \text{false, if the string "}\neg p(t_1,\ldots,t_k)\text{" is in D,} \\ \text{void, in other cases.} \end{cases}$$

**Example 2**. Let $Q=\{q_1, q_2, q_3\}$, $D=\{p(q_1,q_1), p(q_1,q_2), p(q_2, q_2), p(q_3,q_2), p(q_3, q_3)\}$.
Then a system $F_0$ with the only form $\{S=\{\forall x \in S \forall y \in S\ p(x, y)\}$ defines two variants of sets: $M(S)=\{q_1\}$,

$M(S)=\{q_2,q_3\}$.
For the first case $[\forall x \in S \forall y \in S\ p(x, y)] = p(q_1,q_1)$, for the second case
$[\forall x \in S \forall y \in S\ p(x, y)] = p(q_2,q_2) \wedge p(q_3,q_3) \wedge p(p_2,q_3) \wedge p(q_3,q_2)$.

Both are true on the standard interpretation.

**Def. 23.** For an element $q \in Q \cup Q^\lambda$ **the collection of its properties** is called a set of expressions of D containing $q$. This set we will denote as $\varphi(q)$.
For previous example $\varphi(q_1)=\{p(q_1, q_1), p(q_1,q_2)\}$.

**Def. 24. Full collection of properties** of element $q \in Q \cup Q^\lambda$ is defined as a result of next process:
- Each element $q_p$ from arguments of $\varphi(q)$ are replaced with the string $q_p[\varphi(q_p)]$ if $q_p \neq q$;
- Repeat the rule above for each new elements appearing on previous steps.

For an example above, the full collection of properties for $q_1$ is
$q_1[p(q_1,q_1), p(q_1, q_2[p(q_2, q_2), p(q_2,q_3[p(q_3,q_3), p(q_3,q_2)])])]$.

**Def. 25.** The standard interpretation J of a system F we will call **minimal for set $S_a$** if for each such standard interpretation $J_1$ that $Q(J_1)=Q(J)$, $Q^\lambda(J_1)=Q^\lambda(J)$, $D(J_1) \subset D(J)$ and $D(J_1) \neq D(J)$, take place $M(S_a,J_1,F) \neq M(S_a,J,F)$.
Other word, a removing some elements from diagram of a minimal standard interpretation changes the set with name $S_a$.

**Lemma 1.** If for a system $F_1$ minimal for $S_a$ standard interpretation is not minimal for the same set $S_a$ for a system $F_2$, then $F_1$ is not equal to F2 for the set $S_a$.
**Proof.** Let t be an element of D(J) and removing it does not change the $M(Sa,J,F_2)=M(Sa,J_1,F_2)$. Let $J_1$ is made from J by removing t from D(J), then $M(Sa,J,F_2)=M(Sa,J_1,F_2)$ and $M(Sa,J,F_1) \neq M(Sa,J_1,F_1)$, because J is a minimal for $F_1$. If we suppose that $F_2$ is equal to $F_1$ on Sa, then $M(Sa,J_1,F_1)=M(Sa,J_1,F_2) =M(Sa,J,F_2)= M(Sa,J,F_1)$, that contradicts to $M(Sa,J,F_1) \neq M(Sa,J_1,F_1)$. End of proof.

**Def. 25.** Let J be a standard interpretation of F, $\psi_1,\ldots, \psi_k$ be a logical functions. **The secondary for standard interpretation** J is called an interpretation $J^{sec}(J)= (Q \cup Q^\lambda, D^{sec})$, where $D^{sec}$ is the extension of D with expressions "$\psi_i(t_1,\ldots t_p)$" if it is true on D or with the expression "$\neg \psi_i(t_1,\ldots t_p)$" if it is false on D. D' is called a **secondary diagram**.

Let's $\min_{Sa} J^{sec}(J)$ denote a minimal for $S_a$ interpretation $(Q \cup Q^\lambda, D^{min})$ for which $D^{min} \subset D^{sec}$.

**Remark 2.** For each element q replacing logical functions with a disjunctive normal formula with only basic predicates we will got expression as
$(r_i^1(t_i^1) \wedge \ldots \wedge r_i^m(t_i^m)) \vee \ldots \vee (r_j^1(t_j^1) \wedge \ldots \wedge r_j^m(t_j^m))$,
where r is p or $\neg p$. This expression is true if at least for one s $q[r_s^1(t_s^1) \wedge \ldots \wedge r_s^m(t_s^m))]$ is in the diagram D.

All next results of comparison classes are based on a structure of collections of properties of bearer's elements.

One of the basic results of the theory of models in mathematical logic is that names and secondary predicates can be eliminated. So there they are used only for short. Next result shows that if to use a comparison defined in **Def. 4.4.3,** then a hierarchy of names increases the expressive power of system of forms.

**Theorem 1.** Let $G^n$ to be a class of forms with no parameters in set names, $G^{\neg n}$ to be a complimentary ($G^{\neg n} = U - G^n$) set to $G^n$. Then $G^{\neg n} < G^n$.

**Proof**. The truth of $G^{\neg n} \leq G^n$ is come from the fact that by adding a form with a new type of parameter names to any system from $G^{\neg \alpha}$ creates a system from $G^\alpha$ equivalent to original.
Now we will prove that $G^{\neg n} < G^n$ strictly.
Let's $F_0$ be a system of two forms
$S_0 = \{\forall S_x \in S_0 \, \forall y \in S_0 \, \forall z \in S_x \, \forall u \in S_y \, r(z,u)\}$
$S_z = \{\forall x \in S_z \, \forall y \in S_z \, p(x, y, z); z \in S_1\}$

One can see that collection of properties of an element q of a set $S_z$ on a minimum interpretation has a form
$q[\varphi: q \in S_z \in S_0] = q[\varphi: q \in S_z, r(q, q_1^1[\varphi: q_1^1 \in S_{z1} \in S_0]), \ldots, r(q, q_1^1 \in S_{zn1} \in S_0]), \ldots, r(q, q_1^m[\varphi: q_1^m \in S_{zm} \in S_0], \ldots r(q, q^m_{nm} \in S_{zm} \in S_0)])]$, where $\{q_1^j, \ldots, q_{nj}^j\} = M(S_z^j)$ for $S_{zj} \in S_0$; m is a number of names $S_{zj}$ from $S_0$.

Here the record $q[\varphi: q \in S \ldots]$ is for short of the list of properties that determine that $q \in M(S)$.
Replacements them with full records gives a record containing $< h^h$ base predicate r, where $h = \Sigma n_{zj}$.
For a system from $G^{\neg n}$ sizes of collections of elementary properties are less than production of sizes of sets under quantifiers of any formula. Size of collections for elements of these sets also restricts the production of sizes of other forms.
Let's add to the bearer Q new elements and to D such properties that increases each $M(S_{zi})$ ($S_{zi} \in S_0$) independently for $d_{zi} \gg n_{zi}$ elements. Then a size of properties in collection for q will be proportional to $h^h$, where $h = \Sigma d_{zi}$.
For any of systems L from $G^{\neg n}$ the size of properties could not increase more than $\Sigma d_{zi}$. Number of sets used for forms $F_j$ in L is a constant $k_j$, number of form is a constant so number of properties in property collections for any element is not exceed $C_0 * d_{zi}^{C_1}$ for some constants $C_0$ and $C_1$. So selecting $h > C_1$ we may conclude that in $G^{\neg n}$ no system with such properties. Proof is finished.

With the same technique can be proved the
**Theorem 2**. Let $G^r$ is a subclass of systems where each set name $S_a$ has a constant low index *a* and hence number of all set names is a constant. Then $G^r < G^c$.

Let G be a class of all systems. Then next statement is true.

**Theorem 3.** $G^{\neg \beta} < G$.

Proof. We will proof that for system with one β-form
$$S_0 = \{\forall x \in S_0 p(x)\}$$
there is no equivalent in $G^{\neg \beta}$.

Suppose that $S_0$ can be be defined in $G^{\neg \beta}$. Then this form cannot be α type, because sets of properties for its elements for minimum interpretation are empty. It cannot be δ type because $S_0$ does not have any functions in it definition.

Last case is if $S_0$ defined with form γ.

Let $S_b$ be a base set of the definition. If $S_b$ is defined with form of type α then $S_b$ has an elements with empty set of properties and hence assumption that $S_0$ can be defined in class $G^{\neg \beta}$ wrong. If $S_b$ is defined by form γ then consider its base set and so on while all forms from F will be exhausted. Proof is finished.

Similar statement is true for γ type of definitions.

**Theorem 4.** $G^{\neg \gamma} < G$.

**Proof**. Let F is a system with two forms
$S_0 = \{c_1, \ldots, c_k\}$
$S_g = \{S_0, p(x,y), x \in S_g, y \in S_g\}$.

So $F \in G$ and $\neg(F \in G^{\neg \gamma})$.

Suppose the theorem is not true, there is a system $F1 \in G^{\neg \gamma}$ and F1 define the same set S as F. Let's consider interpretation J with a special diagram D constructed with next way. Let $q[\ldots p(q_j, c_i)]$ is an extended representation of element of M(J,F,S), its deepest element has to be from S0.

The deepest element (let's denote it $q_b$) in F1 has to be from S, be an argument of $p(q_a, q_b)$ and be under existential quantification because in other case there is no deepness bigger than summation of sizes of all forms. Deepness for F does not have restriction. Then if $q[\ldots p(q_j, c_i)]$ is an element of set for S in F, then replacing $p(q_j, c_i)$ with true $p(q_j, q)$ – because "$p(q_j, q)$" belongs to a diagram - makes element $q[\ldots p(q_j, q)]$ to belong to S.

Let's add to the diagram D a property $p(q_j, q)$ for each $p(q_j, c_i)$.

Then q will have a loop without elements of $S_0$ and a set of elements from this full collection of properties is a selected set for F1 but not for F. Now if we add copy of elements M(J,F,S)-$\{c_i\}$ and properties copied from $q[\ldots p(q_j, q)]$ with replacing elements to its copy we will get new interpretation $J_w$ where F1 has two sets for S, but F has only one. Hence on interpretation $J_w$ systems F1 and F have a different variants of the selected sets for S. End of proof.

**Remark 3.** The proved theorem states that an induction can't be expressed with a recursion. The way of a proof also shows that a reason is in variants: an induction based on minimum sets, a recursion based on maximum sets.

The request that the selected set has to be maximized has downside discribed in the next statement.

**Theorem 5.** Class $G^{\neg\gamma,\delta}$ has systems with many **agreed** sets but named family with maximum set for some name does not exists.

Proof. Let's consider system with two forms:
$S_1=\{(\forall x_1 \in S_1 p(x_1)) \wedge (\forall y \in S_2 \exists z \in S_1 r_1(z,y))\}$
$S_2=\{(\forall x_2 \in S_2 p(x_2)) \wedge (\forall y \in S_1 \exists z \in S_2 r_2(z,y))\}$.

Let for some interpretation M1 and M2 are subsets in **agreement**. The maximum set for S1 is $M_1'=\{x|p1(x)\}$, but this set can be satisfied with second only if formula
$\forall y \in M_1' \exists z \in S_2 r_2(z,y)$
is true with this $M_1'$. Clearly, that exist interpretations where this formula is not true.

**Theorem 6.** Let's enriched signature of systems with interpreted predicate $x \in S$ and denote its class $G^s$. Original class of systems let's denote G. Then
$G < G^s$.

**Proof.** In systems from G it is impossible to define a complementary set for the set defined with numerated form. In $G^s$ it is obviously possible.

**Discussion.**
The results shown here are just the beginning of studying of nonprocedural definition of subsets. Possible questions for further researches might be new forms of definitions and a **set naming**. So far we assume that names are given independently of the sets elements, but if to make sets names dependent of its elements does it increase the expression power?

### 6. The complexity of algorithms of building defined sets

Results of the previous paragraph show that systems of definitions of named set is a powerful tool for a program specification. The main obstruction to use it as a programming language is a still open satisfiability problem.

But for a wide enough subclasses fast, at least polynomial algorithms exist.

We will see that some additional info about bearers and predicates helps to reduce complexity to polynomial or even leaner dependency.

For instance if for definition
$S=\{\forall x \in S \exists y \in S\ p(x,y)\}$
it is known that p is asymmetric predicate ($p(x,y)= \neg p(y,x)$) , then $M(J,F,S) = \emptyset$ for any F and J.

Another simplification comes from knowledge that for each element x of S exists only one element y of S so that $p(x,y)$ =true.

The leaner algorithm is next.
Let's build a sublist of pairs
$P=\{(x,y)|p(x,y)=true\}$
with the next way. Take a first element of a given finite start data (universe), let it be **a**, and make next step. Find first not selected element **b** that $p(\mathbf{a},\mathbf{b})$=true. If there is no such element, then add pair (**a**, ∅) to P, and **a** is called dead-end element. Otherwise repeat the step. When all elements will be checked, the set of non dead-end elements from P is a set for S.

**Remark 4.** Let's note an important feature of selected set. It was defined as a set which **can't be increased by adding only one element**. The next example shows that it may exist two variants M1 and M2 so that M1 ⊂ M2, but M2 can't be constructed by adding to M1 elements one by one.

The example system contains one form
$$S=\{\forall x \in S \forall y \in S\ p(x,y)\},$$
universe is
$$V=\{a,b,c,d,e,f\},$$
diagram is
$$D=\{p(a,b), p(b,a), p(c,d), p(d,c)\}.$$

Then M1={a,b} and M2={a,b,c,d}.
Clear M3=M1∪{c} or M4=M1∪{d} are not a variant for S.

Let's divide systems of form for subclasses with different number I of form types, number J of quantifications{∀, ∃}, number of their changes K and identify them with vector (I,J,K). So subclass (1,1)=(1,1,0) includes systems with one type of forms and one quantification (0 quantification changes). Algorithms for selection subsets for them are straightforward.
The subclass (1,2,0) consists of forms of next 4 types:
1) $S=\{\forall x \in S \forall y \in S\ p(x,y)\}$
2) $S=\{\forall x \in S \exists y \in S\ p(x,y)\}$
3) $S=\{\exists x \in S \forall y \in S\ p(x,y)\}$
4) $S=\{\exists x \in S \exists y \in S\ p(x,y)\}$.

Farther we will consider a complexity related to the size of data universe.

**Lemma 2.** For subclass (1,2,0) exists a polynomial algorithm of data selection.
**Proof.** For a system of type 1) if a∈M(S) then p(a,a)=true and hence at least one element set can be found for linear time or detect of it absents. If universe V contains such element v, that for already constructed part M for each x∈M
$$p(x,v) \land p(v,x) \land p(x,v) \land p(v,v)=true,$$
then v may be added to M. The part M, which can't be extended such a way, is a variant for S.

**Remark 5.** The algorithm can be generalized for any number r of existential quantifications. Here a selection of the first element **a** for M is made with condition p(**a**,…,**a**)=true and at each step k+1 set $M_k$ is extended with element v if
$$\forall y_1 \in M_k \forall y_2 \in M_k \ldots \forall y_r \in M_k (p(v, y_1,\ldots, y_r) \land p(y_1,v,y_2,\ldots,y_r) \land \ldots \land p(y_1,\ldots,y_r,v)).$$

For a system of type 2) the algorithm creates a subset of universe by removing on each stages elements that is for sure not in M(S). On stage 1 algorithm removes such x∈V that p(x,y)=false for any y∈V. Notes that if V does not have such elements then whole V is a M(S).

On second and on any next step k algorithm remove elements with the same properties but with universe
    $V-\{X_1 \cup \ldots \cup X_{k-1}\}$,
where $X_i$ are elements removed on stage i.
    The process of removing is finished when all elements are removed or no one was removed on current stage. Number of stages does not exceed the size of universe, because at least one element has to be removed.
    Proof of algorithm.
    Let's call element y for which p(x,y)=true, a **supporter** of element x. Then on stage 1 algorithm removes elements that do not have supporter in any subset of universe. If there are no such elements – any element has a supporter - then a whole universe is a set M(S).
On next stages algorithm remove elements which has only supporters that do not have supporters for themselves.
    Let's proof that the rest is M(S).
    Let's suppose there is an element v without support in the rest, meaning its support was removed. But by construction v also have to be removed. The contradiction finishes the proof of current case.

    **Remark 6.** The algorithm may be extended for the form
S={∀x∈S∃y∈S…∃z∈S p(x,y,.., z)} defining supporters for x a set of several elements y,…,z.

    **Remark 7.** Similarly can be build a polynomial algorithm for more general cases of forms with one universal quantification and existential others.

    **Remark 8.** Classes 2) and 4) have a remarkable properties:
  - they have only one variant
  - the complexity of the algorithm above is $n^2$,
  - such type of definitions are close to the induction.

    **Remark 9.** Definitions on the type 2) and 4) can be written as a system of Horn's disjunctions.

    Sets for system of type 3) can be built with extensions of the set for form
    $S_a=\{\forall y \in S_a(p(a, y) \wedge p(a, a))\}$
with a next way. In a current set $M_a = M(S_a)$ find such element b that p(b, x) is true for each x∈$M_a$ and (V-$M_a$) has an element y that p(b, y) is also true. The set $M_a$ is extended with this element y and extension steps are repeated again.
    Systems of type 4) give empty sets if p(x, y) is false for any pair (x, y) ∈V×V or the whole set V if p(x, y)=true for some pair from V.
    End of proof of lemma 4.6.1.

    For class (1.3.0) we already considered cases 1-4, so we have to consider systems
    5) S={∀x∈S∀y∈S∃z∈S p(x, y, z)}
    6) S={∀x∈S∃y∈S∀z∈S p(x, y, z)}.

**Lemma 3.** For interpretations of systems from classes (1.3.1) and (1.3.2) where for each variable under ∃ exists not more than one element to satisfy p(…), exists a polynomial selection algorithm.

**Proof.** Non trivial systems from classes above are

S={∀x∈S∀y∈S∀z∈S p(x, y, z)}

and

S={∀x∈S∃y∈S∀z∈S p(x, y, z)}.

For the first one let's consider interpretations satisfying to lemma: if $p(x,y,z_1)$=true and $p(x,y,z_2)$=true then $z_1=z_2$. Note that if a∈S and if p(a,a,b)=true for an element b then b∈S, because such element is unique, no another element could support a. So any element b that p(x,y,b)=true and x,y are in M(S) also has to be in M(S).

This gives the idea of the algorithm for building selected set.

Take elements **a** and **b** that p(**a,a,b**)=true and add **b** to current set Ma. Then add to Ma any element **c** that p(b,a,c) or p(a,b,c) or (p(b,b,c). If all such elements are already in Ma - stop the building. If there is no such **c** – then Ma is empty and algorithms starts with new **a**. If all **a** were tried then stop – selected set is empty.

The final set Ma obviously satisfies to the form and a complexity of this algorithm is $O(n^3)$.

For a system of second type one of sufficient conditions for existing of a polynomial algorithm is the non emptiness of variants for more strictly form

S1={∀x∈S1∃y∈S1∀z∈V p(x,y,z)} (V is an universe).

Obviously that if M(S1) is not empty, then it satisfy for original form

S={∀x∈S ∃y∈S ∀z∈S p(x,y,z)}.

If to use a predicate p1(x,y)=∀z p(x,y,z) then form

S1={∀x∈S ∃y∈S p1(x,y)}

defines the same variants and belongs to class (1.2.0) which is polynomial.
End of proof.

Another **sufficient condition** is a commutativeness of third and first or second arguments of a predicate p. In this case applicable algorithm for system (1,2, 0) with universal quantifications ∀ only.

**Third sufficient condition** is if an argument under existential quantification is a (Scolem's) function of one or all others arguments. In this case we again may replace 3-D predicate with 2-D and formula has only ∀ quantifications.

**Forth sufficient condition** is if a predicate has separateable variables:

p(x,y,z)=r(x,y)*t(z).

For the case of '*'='∧' let's select such elements z of universe V that t(z) = true. Denote this set of elements as V1. The form 5) is equivalent on V1 to some form of class (1,2) :

S1={∀x∈S1 ∃y∈S1 r(x,y)}

Its building algorithm is polynomial as it was shown earlier.

For the case of '*'='∨' a selected set is a union of sets for

$S_1$={∀z∈$S_1$ t(z)}

and

$S_2$={∀x∈$S_2$∃y∈$S_2$ r(x,y)}.

**Fifth sufficient** condition is if universe V may be factorized for small number of sets $V_1,\ldots,V_r$ (say r independent or logarithmically dependent of the universe size) for which
$$\forall u_1 \in V_j \ \forall u_2 \in V_i \ \forall y \in V \forall z \in V \ (p(u1,y,z) \equiv p(u2,y,z)).$$
In this case

$$\forall x \in S \ p(x, y, z) \equiv p(a_1, y, z) \wedge \ldots \wedge p(a_r, y, z)$$

where $a_1 \in S \cap V_1, \ldots, a_r \in S \cap V_r$ are arbitrary elements. Because r is a constant form 5) is equivalent to form (1,2,0).

Another way to build selected sets is to use simpler form to define **approximation of sets**. For the class 5) form
$$S = \{\forall x \in S \ \exists y \in S \ \forall z \in S \ p(x,y,z)\}$$
Such approximation the form
$$S1 = \{\forall x \in S_1 \exists y \in S_1 \ p(x,y,x) \wedge p(x,y,y)\}.$$
Clear, that M(S)=M(S1). If M(S1) is small then set M(S) may be found by direct checking. Also we may work with smaller universe V=M(S1), where
$$S = \{\forall x \in S \ \exists y \in S \ \forall z \in S \ p(x,y,z); x \in M(S1), y \in M(S1), z \in M(S1)\}.$$

### 7. Fuzzy sets

A natural extension of a set definitions is using fuzzy values from interval [0,1] instead of two-element value {0,1} for belong relations. This extension is used for data mining, objects taxonomy, determine semantics of a text parts.

There are several ways to combine this notion with set definitions.

First is to use fuzzy predicates in set definitions, second – to use fussy quantification, third - to use fuzzy function for predicate of "x is an element of S" used in forms, forth – to use number of elements of sets with some relations as a measure of belonging to the set, fifth and so on may be combinations of these ways.

**Example 4.** Fuzzy predicates are color of the objects – pale red, dark red, or text relation to the subjects – main issue, just mentioned, ….

**Example 5.** Fuzzy quantifications $W^t x \in S \ p(x)$ may be used for an expression like "more than t=10% of elements of S have to have a property p(x)".

**Example 6.** Fuzzy relation $x \in S$ for hierarchical selectors in forms may be useful for selecting elements with a strong relation to some set.

**Example 7.** Fuzzy measure of a weight of element x in the set by counting of elements y of this set having relation p(x,y).

These are new and rich objects for researches. To select the most fruitful direction probably better to start with applications like unstructured text analysis.

## 8. Systems of logical equations for named sets

Let's for each data **d** from universe and each set name $S_i$ map a logical variable $ld_i$. It is true if $d \in S_i$ and false if not.

Then form of definition for $S_i$ generate a system of logical equations

$ld_1 \wedge \ldots \wedge ld_n = $ true.

The solutions of the system are elements for selected sets.

The complexity of this task is not less than complexity of satisfiability problem because the task contains them. But some heuristic algorithm for solving this problem for $O(n2)$ time for 90% of randomly generated systems exists.

## 9. The non-procedural language for business data processing

A data processing can be seen as a step by step simultaneously transformation elements of some named sets of data.

For scientific needs, as we saw in the chapter 2, these sets are linear hyperspaces in multi-dimensional arrays. Business data have more complex structures and are needed another type of descriptions.

In this chapter we will use systems of data set definitions not only for selections but for processing them as well. For this we will use **open systems** of definition when some of sets considered as 'a priory' given sets, not defined with a current system of definitions. Programs are finite sets of open definition systems. One system is declared as a start one. On each next step are applied systems with updated on previous step some of open sets.

So we are using data control instead of direct control as in most of modern languages. Data driven processing have its pro and cons. Main advantage is that adding new functionality is simple – just add one more open system of definitions. It can be done without recompiling or even postponing current run. Debug also local – if input-output is the same then result also did not change. Disadvantage is less usage so far in practice.

### 9.1. Formal systems for non-procedure data processing

From now up we will consider open systems of data sets definitions. Type of names of input sets for system F are denoted I(F), others types of names are denoted O(F). Open system F is an operator from one family M of named sets with names from I(F) to another family of sets G with names from O(F), notation G=F(M).

G may contain new names and elements if F has of forms of type $\delta$.

**Def. 26. Data processing specification** (**DPS**) is a pair (**F**, **S**) where **F** is a finite set of systems, **S** is a set of names for (start) data sets. DPS goes by steps. For any step k>=1 DPS execution builds a family of named sets $F_i(G_{k-1})$ for those $F_i$ whose input is in start data sets $G_0$ and is in $\Delta = \cup_{0 < i < p} F_{jp}(G_{k-1})$ if k>0. The process is finished when for each system some of input sets are empty or no one was updated on previous steps.

Generally the process is **non determinative**; result depends of an applying order of initialized systems.

**Example 8.** (summation). Let's use some interpreted functions:
- **Abit**(S,k) creates non intersected k-element sets arbitrarily selected from S,
- **plus** (Sa) creates x+y for two elements set {x, y}.

Then the DPS G = {$F_1$:  $S_1$={ **plus**($S_a$); $S_a \in$ **Abit**($S_1$, 2) }} calculates sum of data of set $S_1$. $F_1$ divides set $S_1$ to pairs, calculate sum for each of them and return them to set $S_1$. At next steps it again divides $S_1$ to pairs and so on. The process is finished when S1 has less than two elements.

Next two paragraphs shows functional fullness and control power of DPS.

### 9.2. Functional fullness of DPS

Here we will prove that any partial-recursion function (pr-function for short) in Kleene's formalization can be calculated with some DPS.

For variable $x_i$ in pr - formalization corresponds one-element set with a value of $x_i$. Notation f[$S_{x1} \times ... \times S_{xn}$] means that f is applied to the Cartesian product of elements of $S_{x1}, ..., S_{xn}$ and it equal $f(x_1,...,x_n)$.
Then DPS $O_f[S_{x1},..., S_{xn}] \rightarrow S_z$ for V=N, $V^\lambda$=N we will call **representation of function** $f(x_1,..., x_n)[N^n \rightarrow N]$, for start set $Sx1=\{<x_1>\}$, DPS stops if and only if $f(x_1,...,x_n)$ not void and $S_{x1}$ contains a value $m=f(x_1,...,x_n)$.

**Theorem 7.** A class of DPS with V=N, $V^\lambda$=N, where N is a set of natural numbers, with interpreted functions o()≡0, s(x)≡x+1, $t^n_m(x_1,..., x_n) \equiv x_m$ and predicates '=' and '<' contains representations for any pr-function.

**Proof.** We will base on Kleene's formalization of pr-functions with basic functions o(), s(), t() and operators of compositions**,** primitive recursions and minimization. Proof will be done by induction of number k of operators in pr-function.
For k=0, f is one of basic functions and it representations are:
$S_z$={o[$S_x$]},          //return 0 for any argument,
$S_z$={s[$S_x$]},          //return x+1
$S_z$={$t^n_m$[$S_{x1},...,S_{xn}$]}  //return m-th argument.

Let's suppose statement is true for any k≤n and proof that it also true for n+1.

If f constructed with n+1 operators, then there are three possibilities:
1) f is defined with the last operator of composition
$$f(x_1,.., x_n)=g(h_1(...),...h_m()),$$

2) f is defined with the last operator of primitive recursion

$$|f(0,x_1,\ldots,x_n)=g(x_2,\ldots,x_n), \text{ if } y=0,$$
$$|f(y+1,x_1,\ldots,x_n)=h(y+1,x_1,\ldots,x_n), \text{ if } y>0,$$

3) f is defined with the last operator of minimization
$$f(x_1,\ldots,x_n)=\mu(y,g(x_1,\ldots,x_n)).$$

Functions g and h contain less operators than f, hence not more than n operators and by induction assamption they have a representations $O_g$ and $O_h$ in DPS class. Then f in case of composition can be expressed as DPS

$$S_z=Oh(O_{g1}[S_{x1},\ldots,S_{xp}], \ldots O_{gp}[S_{x1},\ldots,S_{xp}]),$$

in case of primitive recursion - as two systems $F_1$ and $F_2$:

$F_1$: $\{S_z=Og[S_{x1},\ldots, S_{xp}], S_i=\{0\}\}$
$F_2$: $\{S_i=\{s([S_i]); \forall v \in S_y \forall u \in S_i(u<v))\}, \ S_z=Oh(S_i, S_z, S_{x1},\ldots, S_{xn})\}$.

Really, the system $F_1$ is applied only once at the beginning, because set it uses will never changed. $F_2$ can be applied only after $F_1$ because at the beginning sets $S_i$ and $S_z$ are empty. But it will be applied on each next steps because $S_i$ will be overwritten and contain number 0,1,2,..., y. Set $S_z$ will get values $f(0,x_1,\ldots,x_n), f(1,x_1,\ldots,x_n),\ldots, f(y,x_1,\ldots, x_n)$.

When a value of element $S_i$ reaches a value y, on next step it will be Ø and $F_2$ can't be applied.

In the case of minimization, DPS consists of two systems $F_1$ and $F_2$:
$F_1=\{ S_i=\{0\}\}$,
$F_2=\{ S_i=\{s[S_i]; \forall y \in S_u(y \neq 0))\}, \ S_u=Og(S_i,S_{x1},\ldots,S_{xn}), S_z=o[S_i]\}$.
$F_1$ will be applied only once at the beginning.

$F_2$ will be applied on each next steps while $S_i \neq Ø$. It stops when $g(i,x_1,\ldots, x_n)$ is not defined and hence $S_u=Ø$, or when y=0 ⇔ $g(i,x_1,\ldots,x_n)=0$ and i is an element of the result set $S_z$.

End of proof .

### 9.3. Data control and Petri net

Petri net is one of best distributed system model combining power, simplicity and clear visual presentation. The target of the paragraph is to show that data control used in DPS is not weaker than Petri nets.

**Petri net** is a bipartite directed multi-graph with vertexes set P∪T (P∩T=Ø), P is a set of **positions**, T is a set of **transitions**. **Marking** is a function µ:[P→N], N – natural numbers. A transition t is called **allowed** if for each incoming position p µ(p)≥k(p,t), where k(p, t) is a number of edges between t and p. **An initialization of transaction** t with marking µ is called a replacing µ with µ': µ'(p)= µ(p)-k(p,t) for input p and µ'(q)= µ(q)+k(t,q) for output q of transaction t.

By introducing for each vertex a set name and constructing for each transaction t a system of definitions it can be modeled Petri net even with **resistant** edges. But inverse is wrong, not each DPS can be modeled with Petri set. The proof is very technical but strait forward. So we omitted it.

References

1. Stanley M. Schwart. The Model Concept: Nonprocedural Programming for Nonprogrammers. Technical Reports (CIS), University of Pennsylvania, 2007.
2. F. Prost. A static Calculus of Dependencies for A-cube. Proceedings, Fifth annual IEEE symposium on Logic in Computer Science, 2000.